# Network Access Control Technology

*Proposition to contain new security challenges*

**Abdelmajid Lakbabi, Ghizlane Orhanou, Said El Hajji**
Laboratoire Mathématiques, Informatique et Applications
Université Mohammed V – Agdal
Faculté des Sciences - Rabat
Morocco
*lakbabi@gmail.com*, ghizlane.orhanou@gmail.com, elhajji@fsr.ac.ma,

**Abstract**

**Traditional products working independently are no longer sufficient, since threats are continually gaining in complexity, diversity and performance; In order to proactively block such threats we need more integrated information security solution. To achieve this objective, we will analyze a real-world security platform, and focus on some key components Like, NAC, Firewall, and IPS/IDS then study their interaction in the perspective to propose a new security posture that coordinate and share security information between different network security components, using a central policy server that will be the NAC server or the PDP (the Policy Decision Point), playing an orchestration role as a central point of control. Finally we will conclude with potential research paths that will impact NAC technology evolution.**

## I. Introduction

Today's networks are not closed entities with well-defined security perimeters; mobile users bring their laptops and mobiles devices in and out of the office. Remote-access users connect from homes and public locations. Business outsourcing requires direct partner access into the internal network. Onsite visitors, vendors, and contractors may need physical access to the internal network to accomplish their work. Even traditional, "in-the-office" workers are subject to threats coming through Internet access, e-mail use, instant messaging, and peer-to-peer (P2P) activities.

Traditional security products acting independently, such as intrusion detection and prevention (IDS/IPS) technology, antivirus measures, and firewalls, are no longer adequate - network traffic is too diverse to rely on these measures. According to a recent Cyber security survey [1], Insider Attacks Are More damaging; Consequences include loss of intellectual property, disclosure of confidential information, violation of privacy laws and loss of money.

In the following section, we will study the Network Access Control technology, its architecture, its components and some top NAC products.

## II. The Network Access Control technology

Network Access control (NAC) mechanism consists basically of two types of assessment:

- User authentication
- Device compliance evaluation

### A. Network Access Control (NAC) architecture

Below, Figure 1 presents the NAC solution overview.

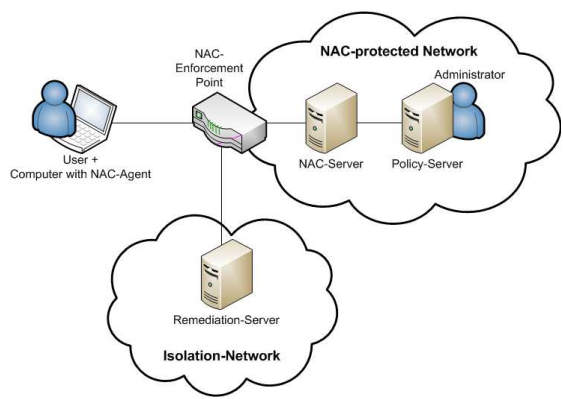

Figure 1: NAC solution overview

This is the process of dynamically provisioning network access for each user and endpoint device. NAC solutions entail authentication (identity), endpoint compliance, remediation, and policy enforcement functions, in the process of validating user identity and the security posture of host devices, before allowing access to the network.

*a) Security products selection Process*

With the idea to select the best security products and tools to build the targeted network security platform, Gartner [2], with a set of technical and commercial criteria for evaluating security products, it can help to approach the most secure solution for each technology layer.

As to NAC solution, Gartner states that Cisco NAC [3] (Network Admission Control) and juniper UAC [4] (Unified Access Control) are the best NAC offer at this moment according to Gartner, as presented below in "Figure2".

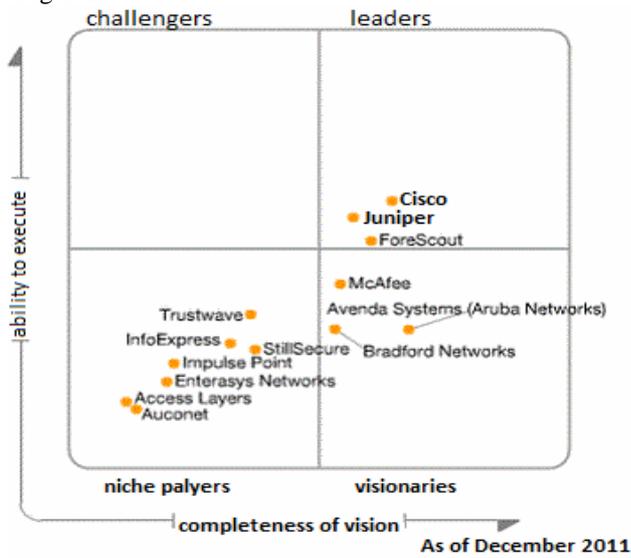

Figure 2: Gartner NAC products classification

In the following subsections, we will compare the two top NAC solutions according to Gartner classification, discuss their respective weaknesses, and then study how NAC can play a fundamental role, to improve network security by extending its capabilities to administer network access requests based on NAC capabilities, and integrating legacy security products, and existing network infrastructure.

*b) Technical description of Cisco and Juniper NAC*

*1. Cisco Network Access Admission overview*

*Cisco NAC mechanism is based on the following process flow as described below in "Figure 3"*

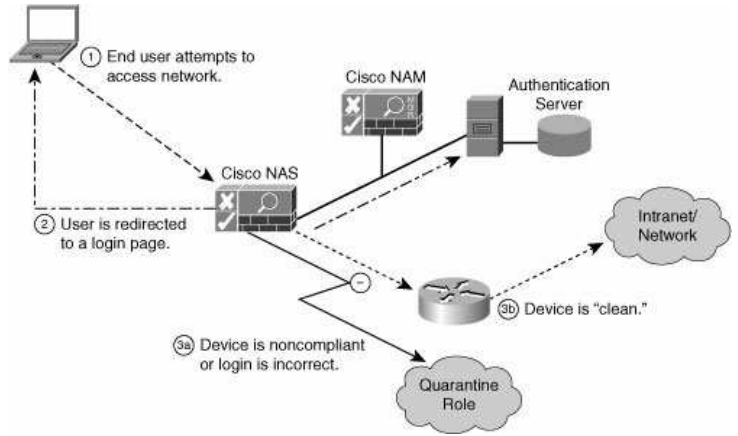

Figure 3: Cisco NAC process flow

Cisco NAC access decision is based on:

Users, their devices, and their roles in the network

Evaluate whether machines are compliant with security policies

Enforce security policies by blocking, isolating, and repairing noncompliant machines

Provide easy and secure guest access

Audit and report whom is on the network

**Enforcement Points** (where the access decision is applied)

- Cisco Switches
- Cisco Routers with NAC modules
- Cisco VPN concentrators

<u>Cisco NAC Weaknesses</u>

- Cisco is ignoring TNC[5] the Trusted Computing's proposed standard
- It is a closed solution that may introduce interoperability issues with third party software and networking equipments
- The OOB (out-of-band) [6] deployment model, requires support for communication between the switch and the Cisco CAM (the Manager need to send and receive SNMP messages to/from Switchs). This is supported only on selected Cisco products.
- Bring security enforcement deeper into the core of the network, but with limited integration with others Cisco network systems, and with no integration with different security products than Cisco.

2. *Juniper UAC overview*

*Juniper NAC mechanism is based on the following process flow as detailed below in "Figure 4"*

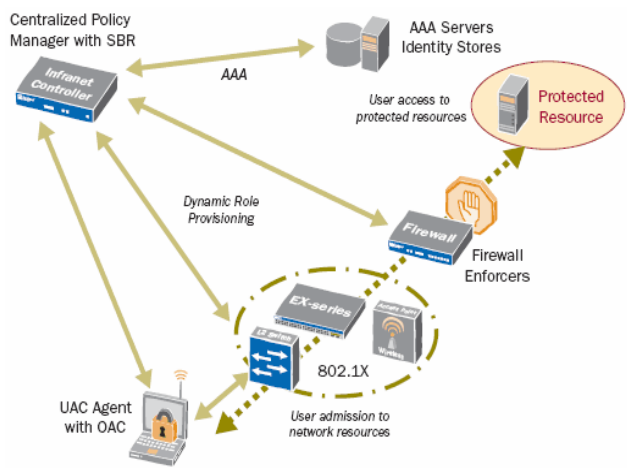

Figure 4: Juniper UAC process flow

Juniper dynamic access control is based on:

User identity

Device security state

Location

**Enforcement Points**

- Policy enforcement provided by EX-series switches and SSG/ISG Firewalls
- IC can push policy name to EX-series switches for dynamic configuration based on user or device
- Policy on EX-series can enforce specific QoS queuing or scheduling policies, VLAN assignment, or any other port configuration parameter

Juniper UAC introduces Coordinated Threat Control with the ability to leverage Juniper's Intrusion Detection & Prevention (IDP) and Unified Threat Management (UTM) products to deliver dynamic network protection, and dynamic User Quarantines as well.

Juniper's UAC enables to leverage the deep packet, application level threat intelligence of Juniper Networks standalone Intrusion Detection and Prevention (IDP) platforms as part of its framework. When a standalone Juniper IDP detects a network threat of a particular type – policies can be configured on several attributes including attack category, attack protocol, attack strings, actions taken, destination or source addresses/ports – it can signal the Infranet Controller, which after receiving the signal and information from the IDP can narrow the threat to a specific user or device; UAC can then implement a configurable policy action, including the following flexible options:

- Quarantining the user (or device) by placing them in a restricted VLAN;
- Changing roles and denying access to certain applications;
- Terminating the user session; or even disabling the user session until an administrator can re-enable it.

*Juniper NAC Weaknesses*

- Juniper's license is restrictive. If a user logs in at two different connections, that will count as two seats instead of one.
- Juniper supports only limited use cases. It does not support routers as an enforcement device.
- It needs an inline firewall for wireless coverage. Juniper's non-802.1x implementation is supported only by the inline firewall at distribution
- Juniper does not provide out-of-box capabilities to manage non-authenticating devices (IP phones, printers, etc.).
- Its auto-remediation capability is limited to basic functions.

c) *Cisco NAC and Juniper UAC security key feautures*

When comparing those two solutions, what is important to retain is not the strengths of each solution by itself, but its ability to interact with others security components in the architecture, using a combination of tactics to provide defense-in-depth to the network; When designing a network access control initiative, it is important to consider interoperability with network infrastructure and existing solutions, NAC initiatives place a high emphasis on the critical combination of security components, and its ability to support these requirements directly determines the global solution's effectiveness; any complete NAC solution will fail if the integration in the existing infrastructure isn't feasible or uses an unsophisticated technique.

As it relates to integration, a NAC implementation is typically best deployed as a solution that creates enforcement points on the existing infrastructure rather than adding extra equipments.

Based on this technical study of the two products, and with the network security collaboration feature in mind, the strategic network security solution will be the Juniper UAC solution, that we will couple with a Juniper IPS to inspect traffic, and take consequent actions against users when their traffic diverge from normal, it enables us to detect that someone is using a non-business critical application, or are exceeding the allowed bandwidth, in such a case, the IPS talks to UAC, then the UAC can take actions like limiting users' bandwidth rates or restricting access.

This integration provides powerful global policy enforcement with centralized management, and ties access control not only to endpoint integrity and user identity, but also to actual traffic through the network.

This feature enables NAC solutions to leverage other security products, like IPS/IDS, as part of the access control deployment, for dynamic threat management, bringing visibility and security enforcement deeper in the network.

### III. OSI Model Enhancement – proposition of a new layer

In today's dynamic computing environment, why have a protection for our network that was built in the past? Today's network security solutions must be able to intelligently recognize friends, collaborators, guests, devices, and suspicious behavior on the network, then take action to prevent security breaches from occurring. An Adaptive Network Security is the key, by integrating and correlating network resources, user, and device information to automate security and IT operations, but to achieve this extended network security policy, extra security information should be considered as we will develop it below using the NAC.

#### a. *THE PROPOSITION TO COUNTER NEW SECURITY CHALLENGES*

With next generation mobile devices, complex networks architectures, new generation of web 2.0 applications like 'Face book', 'twitter', and the challenging network security threats, it becomes necessary to add a new layer to the classic OSI model as shown in "Figure 1", and encourage, even push potentials contributors (Security products Manufacturers & Suppliers, Security designers and Developers) to focus on the identity information in addition to share standardized form of security information and events between their security software and hardware products, acting in different layers of a modified Open Systems Interconnection model that contains eight layers, named "Physical", "Data Link", "Network", "Transport", "Session", "Presentation", "Application", and NAC posture assessment, as we have proposed below in Table 1.

| 8 | **_NAC posture assessment_** | *Standard and Secure Communication Channel over all network security Layers* |
|---|---|---|
| 7 | Application | |
| 6 | Presentation | |
| 5 | Session | |
| 4 | Transport | |
| 3 | Network | |
| 2 | Data Link | |
| 1 | Physical | |

Table 1: The 8 layers networking model

This objective will be materialized by a multi-layer security platform that incorporates the most fundamental security requirements including:

- User/device authentication/posture assessment (AAA layer) to the network security policy.

- A communication channel (Top Down layer) that will warranty secure standardized information sharing among all network security components

#### b. *DEFENSE IN DEPTH CONCEPT*

In fact nowadays, threats are complex and combined, so to fully protect the information during its lifetime, each component of the information processing system, must have its own protection mechanisms. The building up, layering on and overlapping of security measures (defense in depth). The strength of any system is no greater than its weakest link. Using a defense in depth strategy, when one defensive measure fails, there are other defensive measures in place that continue to provide protection.

Controls can be used to form a defense-in-depth strategy. With this approach, defense-in-depth can be conceptualized as distinct layers one on top of the other; More network security can be gained by thinking of it as forming the layers of an onion, with data at the core of the onion, people the next outer layer of the onion, and user identity, network access control, Firewall, IPS and host-based security forming the outermost layers of the onion, as illustrated below in "Figure 5". This perspective is valid and provides valuable insight into the implementation of a good defense-in-depth strategy.

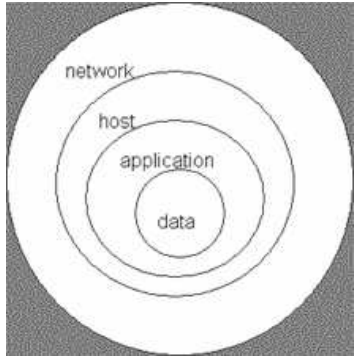

Figure 5: defense-in-depth

To apply this good defense-in-depth strategy, we need to think about security in all layers, and integrate the relevant correlated security information, to make the right network access decision for a given access request, and build an accurate security policy to defeat challenges a network may face.

From the comparison, done in section II, we can easily come to the fact that the NAC is a key solution to control endpoint systems access, based on the user identity and posture assessment, and then play an important role to develop a centralized multilayered security architecture that allows NAC server to act as a policy decision point (PDP). NAC solutions implementation represents an important step towards integrating separate security products in the network, by leveraging functions of directories, AAA servers, network infrastructure devices, and endpoint security software, in the process of dynamically provisioning network access for each user and endpoint device. Most NAC solutions involve authentication (identity), endpoint compliance, remediation, and policy enforcement functions in the process of validating user identity and the security posture of host devices before allowing access to the network. NAC brings identity and compliance awareness into segmentation and access control, its position in the heart of network makes it a central security manager.

SNMP, syslog and proprietary API still play a valuable role with a Security Event Manager (SEM) or similar device like NAC server, to distill the information gathered with theses protocols and feed it into a central database. Also, some flow controllers use SNMP to grant or restrict access

Unfortunately syslog and SNMP both are static and miss the Real-time view of security, that allows products to work together in a coordinated manner to grant access as appropriate while identifying and responding to threats in real time.

Each device reports events but the data is not integrated.

In addition to the new NAC layer, we will analyze the mechanism to share security events via a standard protocol as illustrated below in "figure 6"

Indeed NAC access criteria like location, user device and applications can be used to elaborate a more granular firewall policy rules when using a next generation firewalling technology

IV. Security products and NAC integration

Network security solutions consist of a number of different security standalone products, with each addressing a portion of the overall security needs. So, over network security products undergo dynamic evolution, to achieve a high Multi-layered security platform it is recommended to integer, the more suitable security technologies, in each layer based on the required security features.

Security components basically are integrated one to one using basic protocols and proprietary interfaces, in the perspective to prevent automatically network attacks and take appropriate responsive actions (such as quarantining threatening endpoint devices at the edge of the network where traffic originates).

❖ **NAC server** as the PDP ( Policy Decision Point )

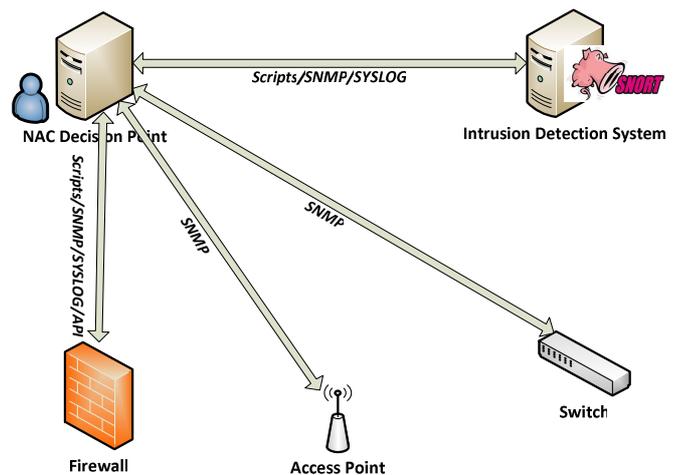

Figure 6: example of NAC Integration

The NAC server, presented above in "Figure 7", represents the central point of decison, and interacts with the rest of the architecture using:

- SNMP ( Simple Network Management Protocol )
- Syslog protocol ( Standard Computer Data Logging)
- Scripts ( Scripting Language )
- API ( Application Programming Interface   )

  ❖ **Next Generation Firewall (NGFW)**

Firewall plays an important role to manage access rules; It can easily be used to dynamically adjust the policy according to NAC system notifications when communication is possible between the two systems.

Since next generation Firewall will filter more than 5 tuple (source IP, protocol, destination IP, source port/local process, destination port/remote process), it becomes necessary to build a firewall policy rules based on the following fields:

*Source:* user/device – That means source IP does not cut it anymore, NGFW should process source based on user authorized roles (AD, LDAP, RADIUS or federated IDs), and should understand if the user is inside the network (LAN), or just coming out of a VPN connection; IT should understand if the user is using an iPad, a blackberry, or a corporate laptop, it can also check if this used device matches the corporate security policy.

Next generation firewall can be integrated with Active Directory and use the user identity as a new type of sources, but they leave posture checks, device identification to NAC / SSL VPN type of solutions.

*Destination:* Destination field should be able to support FQDNs[ since nobody is using a single IP address anymore, and the FQDNs[7] should be dynamically checked bidirectionally – for example voice.google.com may have 100 IP addresses around the globe, and the firewall should block all those addresses with or without name resolution when I write voice.google.com at the destination.  Consequently a NGFW policy rule should contain at least the following fields:

| NAC | | IP Source | IP Destination | protocol | Application | Action |
|---|---|---|---|---|---|---|
| user | device | | | | | |
| **Guest** | **iPAD** | **192.168.1.1** | **www.msn.co m** | **http** | **msn** | **deny** |

Such proposition will have at least two important advantages:

- *Ability to choose the best product in each layer*
- *Security products   integration and events exchange*

Based on that, we will build a multi-layer collaborative network security platform that uses standards protocols and mechanisms to exchange and share security information and events.

  ❖ **Intrusion prevention system (IPS/IDS)**

An IPS (Intrusion Prevention System) is an important component for protecting systems on a network. It is based upon IDS (Intrusion Detection System) with the added component of taking some action, often in real time, to prevent an intrusion once detected by the IDS.

**Protection = Prevention + Detection;** Detecting attacks is a fundamentally different problem than detecting intrusions. Detecting attacks relies on models and patterns of what something bad looks like and then proceed to look for similarities. These systems get their knowledge primarily from external labs and databases.

Detecting intrusions relies on models and patterns of what something good looks like (typically built by base lining normal behavior) and looking for deviations/anomalies. These systems get their knowledge primarily by observing our own networks and systems in use, and then simply have a much higher level of knowledge about the traffic

From this point of view, IDS can regain attention when combined with a policy enforcement system like NAC to take actions against bad sources of malicious traffic; Because IPS systems suffer from some limitations:

- Need to be inline (the traffic pass through) to be able to detect and stop suspicious traffic
- Performance: need to have huge throughput (speed) to keep up with all the inline traffic load especially on the backbone segment

To illustrate this, let's take the following example illustrated in "figure 7" of two DMZ segments where the IPS can't detect attacks inside DMZ2 segment but the IDS can deal with illicit traffic inside DMZ1 since it gets a live copy of the traffic, then send alert to NAC server that terminates the user session; on the other hand the IPS doesn't react to attacks inside DMZ2 (inline mode handle only the traffic that pass from its one interface to another).

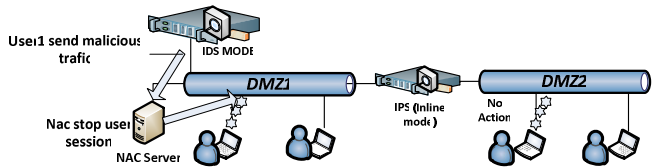

Figure 7: NAC/IDS versus IPS

*Proof of concept and implementation*

To demonstrate the key components and capabilities of NAC solutions, We implemented a Proof-of-Concept plateform in Lab, using PacketFence [8] a Free and Open Source network access control (NAC) solution with a features set including a captive-portal for registration and remediation, centralized wired and wireless management, 802.1X support, layer-2 isolation of problematic devices, and integration with the Snort IDS [9] and the Nessus vulnerability scanner [10].

As an introduction to this implementation phase find below a brief Snort technology description:

Snort is an open source IDS (Intrusion detection system) written by Martin Roesch; Like Tcpdump, Snort uses the libpcap library to capture packets, Snort can be runned in 4 modes:

1. Sniffer mode: snort will read the network traffic and print them to the screen.
2. Packet logger mode: snort will record the network traffic on a file
3. IDS mode: network traffic matching security rules will be recorded (mode used in our tutorial)
4. IPS mode: also known as snort-inline(IPS = Intrusion prevention system)

### V. Conclusion and future work

Securing complex and dynamic network is a big challenge, but the success key is network security collaboration, visibility and the mechanism standardization.

In the network security, there is always a gap, between theory and practice, IF-MAP was experimented in labs, with limited number of devices, and no one can warranty its behavior in a big and complex network, from this perspective it is necessary to deeply focus on this protocol and its performance impact in a real complex network.